\documentclass[doublecol,linenumbers]{epl2} 
\usepackage{amssymb}
\usepackage{graphicx}
\usepackage{dcolumn}
\usepackage{bm}
\usepackage{longtable}
\usepackage{color}
\usepackage{pdflscape}
\usepackage{hyperref}

 \newcommand{\KHC} {KHCO$_3$} 

\newcommand{\KDP}{KH$_2$PO$_4$} 
 \newcommand{\BA} {benzoic acid}

\newcommand{\KDPx} {KH$_{2(1-\rho)}$D$_{2\rho}$PO$_4$}
\newcommand{\Sch} {Schr\"{o}dinger}

\date{\today}

\title{Quantum crystals are at odds with the laws of thermodynamics}
\shorttitle{Title} 

\author{Fran\c{c}ois Fillaux \inst{1}}
\shortauthor{F. Fillaux}

\institute{                    
  \inst{1} Sorbonne Universit\'{e}, CNRS, MONARIS, F-75005, 4 place Jussieu, Paris, F-75005 France.}

\abstract{
Experimental data and the principles of quantum mechanics suggest that a crystal is a condensate of ``wavicles'' enclosed in a box, where ``wavicle'' denotes the indefinite wave-particle status of the microscopic constituents. When it is not perturbed the crystal is in a classical-like state. The wavefunction, the internal energy and the entropy are all of them equal to zero. The thermodynamic temperature is indefinite. Perturbations via energy- and/or momentum-transfer reveal quantum effects. The temperature laws for the heat capacities of some hydrogen-bonded crystals are rationalized with superposition states depending on the quantum temperature which is crystal-dependent. Heat-transfer is a coherent anentropic process. Apart from energy conservation, the other laws of thermodynamics are irrelevant. }

\begin{document}

\maketitle

Thermodynamics deal with heat and temperature of the matter and their relation to energy and work, knowing that heat can be converted to and from other forms of energy, but it cannot be created from nothing nor annihilated. The values of a rather limited number of state-variables at a given moment of time, viz pressure (\textit{P}), volume (\textit{V}), temperature (\textit{T}), internal energy (\textit{E}), entropy (\textit{S})... determine the behaviour of complex systems in the absence of external forces. The state-variables obey empirical laws based on repeated experimental observations. These laws describe and predict a wide range of fundamental phenomena across all fields of natural sciences and technology. It is generally accepted that they emerge from innumerable interactions, which cannot be known in all details, between the elementary constituents of the matter. Statistical thermodynamics linking microscopic and bulk properties account for state-variables of large ensembles of particles. However, irreversible thermalization of an off-equilibrium-ensemble towards an equilibrium state independent of the initial conditions is at odds with time-symmetric dynamics (Loschmidt's paradox). 

Quantum mechanics, as it stands nowadays, is the most plausible theory at the microscopic level and this theory should apply to macroscopic systems just as perfectly as it does to a single entity, as there is no intrinsic limitation against increasing size and complexity. It is legitimate to ask for a quantum account for all conceivable properties of the matter, including those well described by classical thermodynamics \cite{Leggett}. However, the linear formalism of quantum mechanics extrapolated to the macroscopic level leads to dramatic conflicts with our experience of the classical world (\textit{e.g.}, \Sch 's cat) \cite{FSD}. Such conflicts led Bohr \cite{Bohr} to posit a strict separation between the quantum and the classical worlds. Such a separation is not intrinsic to the quantum theory and various \textit{ad hoc} explanations have been proposed, among which decoherence \cite{Zurek2,Schlos}, the stochastic extension of Schr\"{o}dinger's equation \cite{PRC} and spontaneous symmetry breaking \cite{WB1} are notable. 

In the quantum thermodynamics introduced by von Neumann \cite{Neumann} decoherence accounts for the emergence of the laws of thermodynamics from dynamical processes in a vast ensemble of microstates out-of-equilibrium. The key assumption is that the time-evolution is governed by a unitary transformation generated by a global Hamiltonian $\mathcal{H}  = \mathcal{H}_S + \mathcal{H}_B +\mathcal{H}_{SB}$ where $\mathcal{H}_S$ is the system, $\mathcal{H}_B$ is the bath and $\mathcal{H}_{SB}$ is the interaction. This decomposition is not unique but, in any case, $\mathcal{H}_{SB}$ irreversibly turns the pure state representation of $\mathcal{H}_S$ viewed in isolation into a statistical mixture \cite{Zurek2,Schlos,SAB2} and once a quantum superposition has progressed to a mixture of states, no possible measurement can show the effects of quantum interferences. Numerous counterexamples point to the limits of the decoherence scheme. For examples, tunnelling splittings observed in crystals \cite{PH} and gas \cite{KS} reveal that some modes are immune to decoherence. In addition, theoretical models show that decoherence may depend on the preferred decomposition of $\mathcal{H}$ \cite{SM}. In practice, the questions of whether or not coherent states and incoherent mixtures may coexist in the same system and whether or not these states may contribute in different ways to thermodynamical properties are undecidable for the inconceivably large Hilbert-spaces of macroscopic systems. 

For crystals, the most stringent counterfact to decoherence of the bulk is observation, via neutron diffraction, of nuclear quantum-interferences from cryogenic to above room temperature, in addition to regular Bragg-peaks \cite{IF,FCG2,FCG4,FCG5,FCG6,FCou4,FCou5,Fil8}. Such interferences witness to continuous spacetime-translation symmetry and give support to a pure-state representation, referred to as a ``condensate-in-a-box'' (see below) \cite{FCou4,FCou5,Fil8}. In previous works it has been shown that, in addition to quantum interferences, the condensate accounts for the quantum phase-transitions of potassiumdihydrogenphosphate (\KDP) \cite{FCou4} and water \cite{Fil8}, as well as for the proton relaxation rates reported for \KDP\ \cite{FCou4}, \BA\ ($\mathrm{C_6H_5COOH}$) \cite{FCou5} and ordinary water ice I\textit{h} \cite{Fil8}. Here, I present further thermal properties of condensates which compare favourably with published data. It will appear that a condensate is a complete description of a crystal. Heat-transfer is a coherent process and irreversibility follows from breaking the continuous spacetime-translation symmetry of the unperturbed condensate. The quantum temperature of the condensate is different in nature from \textit{T}. Apart from energy conservation, the laws of thermodynamics do not apply to quantum crystals because the state-variables of statistical physics have no counterparts for condensates. 

\section{The condensate-in-a-box} Quantum mechanics is a theory designed purely to predict the probabilities of unpredictable outcomes of quantum measurements. According to Bohr's interpretation \cite{Bohr1,Mermin3}, precisely how the particular result of an individual measurement is brought into being is inherently unknowable and the sort of reality is quantum mechanics about is at odds with the classical notions of waves and particles. An isolated quantum entity cannot be specified in terms of its properties prior to measurement because sometimes it exhibits a wave character, sometimes a particle character, and these states cannot be realized simultaneously. The name ``wavicle'' coined by Eddington \cite{Edd} aimed to denote the indefinite status of quantum entities, but this concept has not been fully appreciated as representative of a physical object of quantum mechanics in place of atoms and waves. Instead it has been perceived as a disturbing renunciation to realism and causality \cite{Mermin3}. 

However, following the definition proposed by Einstein, Podolsky and Rosen \cite{EPR}, wavicles can be regarded as legitimate ``elements of physical reality'' for which indefiniteness is a certitude (probability equal to unity) as strong as definiteness for classical waves and objects is. Indefiniteness is a fundamental character 
of the quantum matter and the mass is the only quantity shared by the classical and quantum theories. Then, a perfect crystal can be thought of as an ensemble of wavicles labeled H, C, O, N... in proportions of the elemental composition. This ensemble is enclosed in physical boundaries, namely the ``box'', in contact with the surroundings at ($T,P$) 
which obeys the laws of statistical physics. 
The boundaries are transparent for probing waves such as photons, neutrons... The fixed number of wavicles in the box is on the order of Avogadro's constant $\mathcal{N}_A$. 

If wavicles were classical entities the crystal at thermal equilibrium would have all characteristics of the microcanonical ensemble, but the indefiniteness of wavicles leads to a drastically different representation that is a superposition of every wave- and particle-states which can be effectively, or in principle, realized via any interaction with the surroundings. In the wave representation, each mode labeled ``\textit{i}'' is represented by a linear combination of eigenfunctions $\psi_{ij}$'s (eigenenergy $h\nu_{ij}$) of the Hamiltonian operator, each with all possible random phases $0 \le \phi_{ij} \le 2\pi$ (mod $2\pi$). Then, the huge numbers of wavicles yields a quasi-continuous and isotropic distribution of $\phi_{ij}$-values and so 
\begin{equation}\label{eq:1}
\Phi = \sum \limits_{i=1}^{N\mathcal{N}_A} \sum \limits_{j=0} ^{j_{imax}} \int\limits _{0}^{2\pi} a_{ij} \psi_{ij} e^{i\phi_{ij}} d\phi_{ij} \equiv 0 . 
\end{equation} 
\textit{N} is the number of degrees of freedom, that is 3 times the sum of the stoichiometric numbers, and $j_{imax}$ is a cutoff preserving the stability of the crystal. The exponentially large Hilbert space of the statistical ensemble shrinks into a single state (entropy $\mathcal{S} \equiv 0$) with continuous spacetime-translation symmetry and a zero-valued eigenfunction. Nuclear positions, energy levels and the time variable are indefinite. The internal energy is $\mathcal{E} \equiv 0$ and the temperature as to the thermodynamic definition $T^{-1} = d\mathcal{S}/d\mathcal{E}$ is indefinite. (An alternative quantum-temperature is introduced below in ``Calorimrtric measurements''.) This condensate is classical-like (say Q-classical) insofar as quantum effects are totally hidden. In fact, a macroscopic ensemble of classical waves with random phases would yield a zero-valued amplitude analogous to (\ref{eq:1}) and it would be impossible to decide whether an unperturbed superposition is either genuinely classical or Q-classical in nature. This overlap of the quantum and classical representations is at variance with Bohr's view of mutually exclusive quantum and classical worlds and it turns out that to claim physical reality for wavicles is not a renunciation to realism. On the contrary, this claim paves the way to an inclusive paradigm for microscopic and macroscopic systems. 

A condensate is a complete representation of quantum and classical properties consistent with Bohr's complementarity principle. On the one hand, quantities measurable without perturbing the condensate, aka ``noninvasive measurements'' \cite{LG}, are definite prior to measurement and the precision of the measure depends on technical limitations exclusively (\textit{e.g.}, mass, density, shape, position of the center-of-mass, electric or magnetic polarization...). Such properties cannot be distinguished from their classical analogues, although they emerge from the quantum. On the other hand, internal observables accessible via invasive measurements involving energy and/or momentum transfer to the condensate (\textit{e.g.}, heat-capacity, energy levels, momentum, nuclear positions...) are indefinite prior to measurements. They are induced by perturbations either at the microscopic level of a single state (see below ``Microscopic events'') or at the macroscopic level of the bath-condensate as a whole (see ``Calorimetric measurements''). 

\section{Microscopic events}

A microscopic event occurs when an impinging wave gets entangled with the condensate. Ideally it consists of an input and an output wave, respectively ($h\nu_0, \mathbf{k}_0$) and ($h\nu_f, \mathbf{k}_f$), the latter being either lost in the environment or captured at a remote detector. The condensate is unaffected by energy and momentum transfer, because these quantities are indefinite before and after the instant of an event. There is no memory of past events imprinted in the condensate and, therefore, there is no decoherence. Measurement-induced states distinctive of a measuring instrument can be inferred from the statistics of energy- and momentum-transfer values effectively measured. 

On the one hand, diffractometers measure elastic-scattering events breaking the continuous space-translation symmetry. The diffraction pattern is the Fourier-transform of the static probability-densities for nonlocal positional parameters in the ``\textit{x}-space'', say $\{x\}$. Time and energy remain indefinite. Quantum interferences observed for mixed isotope crystals \KDPx\ via neutron diffraction \cite{FCou4} is the best hint that the superposition principle holds for wavicles of different masses. 

On the other hand, spectrometers (\textit{e.g.}, those dedicated to absorption of photons or Raman scattering or inelastic neutron-scattering, aka INS) as well as relaxometry instruments (\textit{e.g.}, NMR or quasi-elastic neutron-scattering, aka QENS) detect events breaking the time-translation symmetry. The eigenstates are represented in the $\chi$-space, $\{\chi\}$, defined by the eigenvectors of the nonlocal Hamiltonian operator. Positional parameters and energy remain indefinite in this representation. 

According to Bohr's complementarity, the basis-vectors of $\{\chi\}$ are orthogonal to those of $\{x\}$. These representations are mutually exclusive, even if elastic and inelastic events are not discriminated, as it is the case for many diffractometers. The lack of locally causal links between diffraction and spectroscopy is examplified with \BA\ and \KHC\ \cite{FCou5}. By contrast, conflated $\{\chi\}$ and $\{x\}$ is a tenet of semiclassical theories of many-body-syems concerned with causal-correlations between energy and nuclear positions, \textit{e.g.}, within the framework of the Born-Oppenheimer separation \cite{CB,TC}. Conflation not only is at odds with complementarity, but it is also in contradiction with Bell's theorem whereby quantum mechanics is incompatible with any theory based on local hidden-variables and preserving the postulate that an isolated system can be given a description in its own right \cite{Leggett,Mermin3}. As a consequence, there is no logical contact at all between the statistics of outcomes of quantum measurements and quantized dynamics of classical-nuclei moving across a preexisting potential-energy-surface. 

\section{\label{sec:3} Calorimetric measurements}  A crystal mounted in a thermostat is in contact with an inert gas at controlled temperature and pressure (the heat-bath) and time-averaged collisions at the boundaries lead to thermal equilibrium via heat-transfer. On the very long time-scale of calorimetric measurements, the spectrum of the bath, the details of the microscopic events and the unpredictability of outcomes are overlooked. All possible events and outcomes are certain to happen. Determinism emerges. Repeated measurements of the same system give the same result to within technical precision, exactly in the same way as it is the case for a classical object. 

The indefiniteness of the thermodynamic temperature suggests that the temperature of a condensate is not a statistical average. It is a quantum variable, say the Q-temperature $\Theta$. The crux of the matter is how to rationalize heat-transfer breaking the time-translation symmetry at the macroscopic level, knowing that a condensate is immune to decoherence at the microscopic level. This dichotomy suggests that heat-transfer proceeds via those particular eigenstates involved in phase-transitions breaking the spacetime-translation symmetry of the bulk. This is exemplified by the calorimetric studies of hydrogen-bonded crystals presented below, namely ice I\textit{h} and \KDP. It transpires that the Q-temperatures are effectively different from \textit{T} and largely depend on the crystal-phases, in marked contrast with the universal character of the thermodynamic temperature. 

Ordinary water ice I\textit{h} is an archetypal quantum-crystal \cite{Fil8}. The measured heat-capacity at the temperature of fusion $T_F$, viz $C_{PI}(T_F) \approx 1.01\times\frac{9}{2}\mathcal{R}$, is representative of an ensemble of $3\mathcal{N}_A$ disentangled wavicles. The condensate $|\Psi_{FI} \rangle$ at $T_F$ is, therefore, a pertinent candidate as the main thoroughfare for heat-transfer. As the temperature of ice is lowered, $C_{PI}(T)$ decreases quasi-linearly to zero at $T\lessapprox 7$ K \cite{MK,SLL}. This is at odds with statistical models based upon the crystal density-of-states (\textit{e.g.}, Debye's model). This suggests that the condensate at $T < T_F$ is a linear combination of states including $|\Psi_{FI}\rangle$.  

At $T_F$ the crystal and the liquid are in equilibrium. The heat-capacity of liquid water, viz $C_{Pw} = 9\mathcal{R}$, is distinctive of H$_2$O-entities in the equipartition regime. The so-called ``proton-stretching-modes'' of these entities are eigenstates of asymmetric double-well operators which can be treated as two-level systems. The asymmetry $h\nu_1$ is the energy difference between the state largely localized in the upper well, \textit{e.g.}, $|1_R\rangle$ in the $R$-well, and that largely localized in the lower well, $|0_L\rangle$ in the $L$-well, so that $h\nu_R - h\nu_L = h\nu_{1}$. For the opposite asymmetry, $h\nu_L - h\nu_R = h\nu_{1}$. The same scheme holds for several hydrogen-bonded crystals \cite{FCou4,FCou5}.

Spectroscopic studies suggest that asymmetric double-wells in water at $T_F$ are in equilibrium with pairs of entangled double-wells with opposite asymmetry in ice. This is attested by tunnelling splittings due to the tiny delocalizations of the wavefunctions: $|0_{\pm }\rangle = 2^{-1/2}[|0_{L}\rangle \pm |0_{R}\rangle]$; $|1_{\pm} \rangle = 2^{-1/2}[|1_{L}\rangle \pm |1_{R}\rangle]$; $h\nu_{0\pm } = h\nu_{1\pm } = h\nu_{t}$. 
(The states $|1_{\pm} \rangle$ are distinctive of a superposition of asymmetric double-wells. There is no counterpart for a genuine symmetric double-well of similar shape \cite{FCou4}.) 

At the macroscopic level, fusion occurs when the thermal energy of water is equal to 7 times the eigenenergy of the symmetric state $|{1+}\rangle$: $9\mathcal{R} T_{F} = 7\mathcal{N}_A h(\nu_{1} - \nu_t/2)$ \cite{Note3}. The condensate is $|\Psi_{FI}\rangle = |\Psi_{1+}\rangle ^{\otimes 7}|\Phi'\rangle$, where $|\Psi_{1+}\rangle ^{\otimes 7} = \mathcal{N}_A |1+\rangle ^{\otimes 7}$ and $|\Phi'\rangle$ is the condensate (\ref{eq:1}) for modes others than H-stretching. $|\Psi_{FI}\rangle$ is accompanied by a condensate of tunnelling state $|\Phi_t\rangle = |\Psi_{0-} \rangle^{\otimes7} |\Phi'\rangle$, which has no counterpart in water. The eigenenergy is $E_t = 7\mathcal{R}T_t$, where $T_t= (E_{0-} - E_{0+})/k_B \approx 1$ K. This state is not a thoroughfare for heat-transfer. 

Ice in thermal equilibrium with a heat-bath at $T_F \ge T \ge 7 T_t$ can be represented by:  
\begin{equation}\label{eq:2}\begin{array}{rcl}
|\Phi\rangle_{T} & = & (\alpha |\Psi_{1+}\rangle^{\otimes\aleph} + \beta |\Psi_t\rangle^{\otimes\aleph} ) |\Phi'\rangle;\\ 
|\alpha|^2 & = & \displaystyle{\frac{T -\aleph T_t}{T_F-\aleph T_t}}; \\
|\beta|^2 & = & \displaystyle{\frac{T_F - T}{T_F-\aleph T_t}}; 
\end{array}\end{equation}
where $\aleph =7$ is the dimensionality of the phase transition. $\Theta = T - 7T_t$ is the Q-temperature of the condensate and $\mathcal{V}(\Theta) = |\alpha|^2 9\mathcal{R}T_F + |\beta|^2 \aleph \mathcal{R}T_t$ is the thermal potential-energy. At $T \le \aleph T_t$, $|\Phi\rangle_T$ is forbidden. 

The temperature law for the heat-capacity can be written as: 
\begin{equation}\label{eq:4}\begin{array}{rcll}
C_{PI}(T) & = & \displaystyle{ \frac{9}{2} \mathcal{R}\frac{T - \aleph T_t}{T_F- \aleph T_t};} & \aleph T_t \le T \le T_F; \\
C_{PI}(T) & = & 0; & T \le \aleph T_t. \\
\end{array}
\end{equation}
This law is in full agreement with observations. The discontinuity unveils a second-order phase-transition to a perfect thermal-insulator below 7 K. This transition should occur at a much lower temperature for deuterated ice. 

\KDP\ is also a quantum crystal. Diffraction reveals a network of hydrogen bonds and spectroscopy accords with double-well operators similar to those of water \cite{FCou4}. This crystal undergoes a first-order ferroelectric $\longleftrightarrow$ dielectric transition at $T_0 \approx 122$ K. Upon heating the ferroelectric-phase at $T_0$, the asymmetric double-wells turn into a superposition of double-wells with opposite asymmetry in the dielectric-phase, or vice versa upon cooling. The transition occurs when $12\mathcal{R}T_{0} = 4\mathcal{N}_0 h(\nu_{1} + \nu_{t}/2)$. 
Here, $h\nu_t$ is the tunnelling splitting in the dielectric phase which vanishes in the ferroelectric phase. By analogy with ice, the heat-capacity of \KDP\ can be written as: 
\begin{equation}\label{eq:5}\begin{array}{rcll}
C_{PK}(T) & = & \displaystyle{ 12\mathcal{R}\frac{T}{T_0};} & T \le T_0; \\
C_{PK}(T) & = & \displaystyle{ 12\mathcal{R};} & T_0 \le T \ll T_d. \\
\end{array}
\end{equation}
Below $T_0$, $\Theta = T$. Above $T_0$, the state $|\Psi_0 \rangle$ involved in the phase-transition is the ground-state of the dielectric phase. $C_{PK}$ is a constant, presumably as long as \textit{T} is far below the decomposition limit $T_d$. The Q-temperature is $\Theta = T-T_0$. 

Eq. (\ref{eq:5}) is in agreement with $C_{PK} \approx 1.002\times 12\mathcal{R}$ effectively measured nearby $T_0$, but further confrontation with temperature effects is hampered by the lack of extensive data \cite{FCou4}. Nevertheless, $\Theta$ can be confronted with relaxometry measurements. 

\section{Relaxometry measurements} At the microscopic level, every off-resonance energy-transfer event ($\delta\nu = \nu_0 - \nu_f \approx 0$) realizes one state among all possible states with random $a_{ij}$'s and eigenenergy $\approx \mathcal{V}(\Theta)$:  
\begin{equation}\label{eq:3}\begin{array}{l}
|\psi \rangle = (N\mathcal{N}_A)^{-1/2} \sum \limits_{i=1}^{N\mathcal{N}_A} \sum \limits_{j=0} ^{j_{imax}} a_{ij} |\psi_{ij} \rangle ;\\ 
(N\mathcal{N}_A)^{-1} \sum \limits_{i=1}^{N\mathcal{N}_A} \sum \limits_{j=0} ^{j_{imax}} a_{ij} a_{ij}^* = 1 ;\\ 
(N\mathcal{N}_A)^{-1} \sum_{ij} a_{ij} a_{ij}^* h\nu_{ij} = \mathcal{V}(\Theta) + \mathcal{O}(h\delta \nu). \\
\end{array}\end{equation} 
For non-integrable modes the statistics of outcomes compare favourably with Boltzmann's law:  
\begin{equation}\label{eq:6}\begin{array}{rcl}
\overline{a_{ij} a_{ij}^*} & \approx & \displaystyle{\exp - \frac{h\nu_{ij}}{k_B\Theta}}. \\
\end{array}\end{equation} 
For \KDP, $\Theta = T$ at $T < T_0$ and $\Theta = T-T_0$ at $T>T_0$ \cite{FCou4}, in accordance with (\ref{eq:5}). For \BA\ there is no phase-transition and $\Theta = T$ from cryogenic to room temperature \cite{FCou5}. 

Equation (\ref{eq:6}) is not representative of a statistical ensemble. It simply means that all possible states (\ref{eq:3}) have equal probabilities to be realized (see below ``Thermalization''). $\Theta$ and $\mathcal{V}(\Theta)$ are preexisting ``elements of a physical reality'' outside the bounds of statistical physics, which are distinctive of a particular crystal-phase in thermal equilibrium with a heat-bath at \textit{T}. 

\section{Thermalization} In statistical physics, thermalization is the irreversible dynamical process of equilibration via heat-transfer of an isolated system initially out-of-equilibrium towards a state which is indistinguishable from the microcanonical ensemble. However, Loschmidt's paradox points out that irreversibility is incompatible with time-reversal symmetry. Various hypotheses aim to circumvent this contradiction. 

In classical physics, the axiom of causality (\textit{i.e.}, causes precede effects) is invoked to break the time-reversal symmetry. Dynamical systems prepared in an off-equilibrium state evolve necessarily forward in time and the 
fluctuation theorem gives the probability that the entropy will evolve towards the maximum value corresponding to the microcanonical ensemble. At the microscopic level, the evolution is depicted in phase-space as mode-trajectories over a constant energy surface divided into many domains of equal areas. If the initial conditions are fixed, excluding any point located on a periodic orbit of an integrable mode, the ergodic hypothesis states that the system will be equally likely to be in any one of these domains after a certain finite time. Alternatively, for a range of initial conditions, classical chaos implies that, if the measurement time is fixed but not too early, the system will once again be equally likely to be in any one of the domains. The rule that equal-area domains are equally likely is equivalent to maximum entropy \cite{Srednicki}. 

Alternatively, statistical quantum thermodynamics is concerned with a large number of virtual copies of a system considered all at once, each copy representing a possible microstate obeying Schr\"{o}dinger's equation. Thermalization is impossible in principle, insofar as the time evolution is unitary and the spectrum is discrete \cite{Srednicki,Deutsch,RDO,EFG,Reimann2}. Thermalizatiion requires some \textit{ad hoc} hypothesis among which the ``eigenstate thermalization hypothesis'' (ETH) \cite{Srednicki,Deutsch} aims to explain when and why an isolated system initially prepared in a far-from-equilibrium state can evolve towards a state which can be accurately described using statistical laws at equilibrium. However this hypothesis is not deduced from general theoretical arguments \cite{EFG}, it is neither necessary nor sufficient \cite{Reimann2} and it does not hold for integrable modes. 

In contrast with classical or quantum thermodynamics, the condensate framework is self-contained. It does not require any addition to the quantum theory in order to account for equilibration of a perturbed crystal. 

(i) In calorimetric measurements, an incremental $\delta T$ of the bath at time $t_0$ set the crystal-bath out-off-equilibrium and $\delta Q$ is the heat necessary to relax the system towards equilibrium at $T + \delta T$. Irreversibility stems from the indefiniteness of the time-variable prior to $t_0$ and the arrow of time is unequivocally oriented towards the future. 

(ii) Both ergodicity and chaos are replaced with the quasi-continuous distribution of $\phi_{ij}$'s (\ref{eq:1}) with no exception for integrable modes. 

(iii) Thermalization is a coherent process excluding dynamical equilibration of preexisting eigenstates obeying \Sch 's equation at the microscopic level. Calorimetric measurements do not provide any insight into what happens in the course of time, but the quantum theory states that all possible intermediates are certain to be realized at random. Therefore, it is irrelevant to put forward any particular ``thermalization process'' among a quasi-infinite number of such processes. 

\section{The laws of thermodynamics}  These laws are designed for external observers concerned with macroscopic many-body-systems perceived as classical objects. They rest on the assumption that all state of the matter depend on the same variables within the framework of statistical physics. However, quantum crystals are unquestionably outside the bounds of this framework. 

Thermal equilibrium is a tenet of thermodynamics. By definition, two systems linked by a diathermal junction are said in thermal equilibrium if they do not change over time, so that the net heat-flux between the two systems is equal to zero. For condensates the notion of statistical equilibrium, namely nothing happens on average although 
each microstate evolves with time, is replaced with continuous spacetime-translation symmetry whereby nothing happens in principle. Direct heat-exchange between quantum crystals in contact with each other is precluded by the zero-valued wavefunctions and the lack of decoherence. 

The zeroth law states: \textit{If two systems are in equilibrium with some third body, then they are also in equilibrium with each other}. This law sounds trivial, since the three systems do not change with time, but it is commonly regarded as a necessary condition for definition of the temperature scale such that systems in equilibrium are at the same temperature, whereas when they are off-equilibrium heat flows irreversibly from the hotter to the colder. 

This law is irrelevant for quantum crystals because the $\Theta$ is not a universal state-variable. It depends on the chemical composition of a crystal and its phase-diagram. Two different crystals in thermal equilibrium with a common bath are not necessarily at the same temperature. However, transitivity holds for two distinct baths in equilibrium with the same crystal. They are effectively at the same temperature, but it is not necessarily that of the crystal. If the two baths are at different temperatures, heat can flow from the hotter to the colder bath through the crystal out-of-equilibrium. 

The first law states: \textit{The internal energy of an isolated system is a constant.} For a quantum crystal with continuous time-translation symmetry, energy conservation is a theorem (Noether's theorem), rather than a law. The internal energy $\mathcal{E} \equiv 0$ is invariant and the thermal potential $\mathcal{V}(\Theta)$ is a constant. 

The second and third laws deal with entropy. The second law breaks the time-reversal symmetry: \textit{The total entropy of a system and its surroundings undergoing heat-transfer can never decrease over time}. The third law states: \textit{As a system approaches asymptotically absolute-zero, the entropy approaches a minimum value, that is zero for an ideal crystal. In addition, it is impossible to reduce the entropy of a system to its absolute-zero value in a finite number of operations}. These laws are irrelevant for condensates. They should be rephrased as theorems: 

(1) \textit{The arrow of time induced by heat-transfer is oriented towards the future}; 

(2) \textit{The zero-valued entropy is invariant}. 

\section{Conclusion} Wavicles are physical objects of the quantum theory. This claim applied to many quantum-body systems opens up the vista of quantum condensation at the macroscopic level and emergence of a Q-classical state without discontinuity of, or addition to, the quantum theory. The wavefunction, the entropy and the internal energy are all of them equal to zero. The thermodynamic temperature is indefinite. When it is not perturbed this state cannot be distinguished from a classical object viewed as a continuous and static medium. 

A condensate fulfills all rational demands for a complete description of the statistics of outcomes of measurements. At the macroscopic level heat-transfer is a coherent anentropic process and the quantum temperature is crystal dependent. At the microscopic level, quantum superposition and continuous spacetime-translation symmetry replace ergodicity, chaos and time-reversal symmetry. Lattices or phonons are measurement-induced representations of quantum crystals. Microcanonical average arises from the statistics of unpredictable microscopic events. 

In statistical physics state-variables are universal. On the contrary, there is no universal scale for the Q-temperature. Apart from energy conservation, the other laws of thermodynamics are irrelevant for quantum crystals. 

The present work deals with hydrogen-bonded crystals only. There is every reason to suppose that the framework can be applied to other classes of crystals but this is an open question. 


\bibliographystyle{eplbib} 


\end{document}